\documentclass[pdflatex,sn-mathphys-num,iicol]{sn-jnl}

\usepackage{graphicx}
\usepackage{bm}
\usepackage{amsmath,amssymb,amsfonts}
\usepackage{booktabs}
\usepackage{array}
\usepackage{tabularx}
\usepackage{placeins}
\usepackage{etoolbox}
\usepackage{microtype}
\usepackage{xurl}
\apptocmd{\thebibliography}{\raggedright}{}{}
\usepackage{capt-of}

\newcolumntype{L}{>{\raggedright\arraybackslash}X}

\newcommand{\AUC}{\operatorname{AUC}}

\makeatletter
\patchcmd{\email}{\textcolor{blue}{#1};\ }{\textcolor{blue}{#1}\ }{}{}
\patchcmd{\email}{\textcolor{blue}{#1};\ }{\textcolor{blue}{#1}\ }{}{}
\patchcmd{\maketitle}{Corresponding author(s). E-mail(s):}{Corresponding author. E-mail}{}{}
\patchcmd{\maketitle}{Contributing authors:\ }{Contributing authors\ }{}{}
\renewcommand{\keywords}[1]{\ifx#1\empty\else\def\@keywords{\par\addvspace{10pt}{\keywordfont{\bfseries\keywordname} #1\par}}\fi}
\makeatother

\begin{document}

\title[Robustness-based laser-locking selection]{Robustness-based selection of laser-locking architectures for time-delay interferometry in space-based gravitational-wave detectors}

\author[1]{\fnm{Hao-Kang} \sur{Chen}}

\author*[2]{\fnm{Pan-Pan} \sur{Wang}}
\email{ppwang@cqu.edu.cn}

\author[1]{\fnm{Cheng-Gang} \sur{Shao}}

\affil[1]{\orgdiv{National Gravitation Laboratory, MOE Key Laboratory of Fundamental Physical Quantities Measurement, and School of Physics}, \orgname{Huazhong University of Science and Technology}, \orgaddress{\city{Wuhan}, \postcode{430074}, \country{People's Republic of China}}}

\affil*[2]{\orgdiv{College of Physics}, \orgname{Chongqing University}, \orgaddress{\city{Chongqing}, \postcode{401331}, \country{China}}}

\abstract{Space-based gravitational-wave detectors compare laser phases exchanged between distant spacecraft. Offset phase locking keeps beatnotes in the phasemeter band, while time-delay interferometry suppresses laser frequency noise. A disturbance of one reference can degrade the functions that depend on it, making robustness an additional criterion for choosing a phase-reference architecture. We introduce a quantitative comparison based on established cascade equations and percolation measures, with directed reference graphs and Erd\H{o}s--R\'enyi support layers. Under uniform disturbances, the robustness score is determined by the reference-depth distribution. Under nonuniform disturbances and partial dependence, the connected-support criterion can select a different configuration from both the published depth and local-lock criteria and a reference-path baseline. The phase-locking comparisons favor configurations that limit accumulated degradation and supply downstream lasers from references retaining greater connected support. Paired finite-network calculations support these differences in preference. In a separate comparison of optical-input support for time-delay interferometry, requiring connected support on all spacecraft gives a different preference from requiring it on at least one spacecraft. These findings identify disturbance location, reference dependence and the required extent of support as factors that can change architecture selection. Within the stated statistical model, robustness rankings can complement phase-locking feasibility and interferometric performance when prioritizing configurations for space-based gravitational-wave detectors.
}
\maketitle
\section{Introduction}\label{sec:introduction}

Space-based interferometers extend gravitational-wave observations to the millihertz band through laser links between freely falling spacecraft. The Laser Interferometer Space Antenna, LISA, uses three spacecraft separated by approximately 2.5 million kilometres \cite{AmaroSeoane2017LISA}. Massive black-hole binaries, compact-object inspirals, and Galactic binaries motivate long observations with stable phase measurements \cite{AmaroSeoane2023LISA}. The mission design combines these science objectives with requirements on the optical measurement chain \cite{Colpi2024LISADefinition}. TianQin offers a complementary realization of millihertz detection \cite{Luo2016TianQin}, with fundamental-physics and cosmology objectives developed in a comprehensive white paper \cite{Luo2026TianQinWhitePaper}. Taiji targets related observations with another space interferometer \cite{HuWu2017Taiji}.

A space interferometer obtains time series of laser phase comparisons between distant spacecraft along long, time-varying optical arms. Each one-way phase measurement contains a gravitational-wave contribution, laser frequency noise and secondary instrumental noise. These contributions are evaluated at the emission and reception times set by the unequal light-travel delays. Laser frequency noise exceeds the gravitational-wave contribution by many orders of magnitude. Its suppression therefore forms an essential part of the measurement architecture. Time-delay interferometry, abbreviated TDI, cancels laser noise through delayed linear combinations of the measured phases \cite{Estabrook2000TDI}. Unequal arm lengths determine the cancellation conditions \cite{Tinto2002TDI}, and implementation requires consistent measurements and delays \cite{TintoShaddock2003Impl}. For moving spacecraft, the delay operators retain their time order in second-generation combinations \cite{Tinto2004Moving}. Offset phase locking controls slave lasers at prescribed frequency offsets and keeps selected beatnotes in the phasemeter band \cite{Heinzel2024FrequencyPlanning}. The measured phases then enter the delayed combinations used for laser-noise suppression \cite{TintoDhurandhar2021Review}.

Both operations admit several configurations. For a fixed primary laser, the catalogue of Heinzel et al. contains six non-swap and twelve frequency-swap reference trees \cite{Heinzel2025Catalog}. Their paths and local-lock counts provide criteria for reference transfer and frequency planning. Their residual phase correlations also affect TDI noise spectra \cite{Staab2024LaserResiduals}. TDI combinations have distinct gravitational-wave responses and noise sensitivities \cite{Wang2023GeometricTDISensitivity}. They also differ in the estimation of tilt-to-length coupling coefficients \cite{Chen2025TTLTDICombinations}. Their response functions enter polarization constraints \cite{JiangZhang2026EPJCPolarization}, analytical sensitivity calculations \cite{Zhang2026EPJCSensitivity}, and dark-matter searches \cite{Liu2026EPJCTDICombinations}. These studies compare configurations using the measurement and science requirements associated with each application.

Robustness introduces a further comparison. A disturbed phase reference can degrade the lasers that use it, while the usefulness of a TDI measurement depends on its required optical inputs. Existing work already addresses robust frequency transfer through reference paths \cite{Heinzel2025Catalog}, robust signal extraction using selected TDI combinations \cite{Wang2024MinimalNullTDI}, and noise characterization with different observables \cite{Wang2024RobustNoiseTDI}. Here the comparison concerns the final fraction of qualified functional support after a specified disturbance propagates through an architecture. The dominant mechanisms differ between reference transfer and optical-input support, so their graphs are defined separately.

Percolation in interdependent networks supplies an established mathematical description of repeated dependency and connectivity degradation \cite{Buldyrev2010NatureCascade}. Partial dependence changes the transition behavior \cite{Parshani2010CouplingPRL}. Generating functions specify the connected support response of an individual network \cite{Newman2001Generating}, and the corresponding equations for a network of networks determine its coupled stationary state \cite{Gao2013Framework}. Here the physical reference relationships determine the macrostructure, while Erd\H{o}s--R\'enyi connectivity, abbreviated ER, and the giant-component support rule define the internal statistical model. Applying this framework to a reference tree gives an analytic reduction under uniform initial retention. Assigning different initial usable fractions then tests whether the same architecture remains preferred under a spatially localized disturbance.

Published locking configurations are compared with ER and path responses under identical initial fractions and dependency assignments. A non-swap example distinguishes catalogue choice N1, path choice N6, and ER choice N3 through the support available from competing providers. The complete grid retains ties, coincident choices, and the full-dependency control. Finite network calculations test the ensemble equations and cascade stages. Separate TDI-support calculations check optical-input direction, graph equivalence, and transition locations. These comparisons identify the disturbance and dependency conditions under which internal connectivity changes architecture selection.

\section{Phase references and TDI measurements}\label{sec:physical_principles}
\subsection{Spacecraft, lasers, and phase measurements}\label{subsec:one_way}

The modeled constellation has three spacecraft. Each carries two lasers aimed at the remote spacecraft, optical benches, test masses, photoreceivers, phasemeter channels, timing hardware, and data handling units \cite{BayleHartwig2023Unified}. The laser $L_{ij}$ is on spacecraft $i$ and transmits toward spacecraft $j$. Indices $i,j,k$ label spacecraft in this section, while $u,v$ label individual lasers. Measurement $s_{ij}$ is formed on spacecraft $j$ by interfering light received from spacecraft $i$ with the assigned local oscillator. This sender-first convention reverses the receiver-first ordering used for catalogue beatnotes.

Let $\phi_u$ be the phase variation of laser $u$ after subtraction of the common nominal optical carrier. For laser $L_{ij}$, this phase is written $\phi_{ij}$. The deterministic carrier-propagation contribution along the nominal orbit is subtracted from $s_{ij}$. All phase terms are expressed in radians. The simplified phase-measurement model follows the one-way relation used in TDI \cite{TintoDhurandhar2021Review},
\begin{equation}
 s_{ij}(t)=\mathcal D_{ij}\phi_{ij}(t)-\phi_\ell(t)+H_{ij}(t)+N_{ij}(t).
 \label{eq:one_way_phase}
\end{equation}
Here $H_{ij}$ is the gravitational-wave contribution, $N_{ij}$ collects the secondary instrumental contributions, and $\ell$ labels the receiving local oscillator assigned to this link. The light-travel delay is
\begin{equation}
 \mathcal D_{ij}a(t)=a[t-\tau_{ij}(t)],\qquad \tau_{ij}(t)=\frac{\mathcal L_{ij}(t)}{c},
 \label{eq:delay_operator}
\end{equation}
where $\mathcal L_{ij}$ is the nominal optical arm length and $c$ is the speed of light. Time-dependent delays retain their order in TDI \cite{TintoDhurandhar2021Review}.

The local oscillator is laser $L_{ji}$ for non-swap hardware and $L_{jk}$ with $k\ne i,j$ for frequency-swap hardware. The latter uses the local laser aimed along the other arm \cite{Heinzel2025Catalog}. The local-oscillator assignments and optical layout remain fixed within each hardware family throughout each comparison.

\subsection{Laser offset phase locking}\label{subsec:phase_locking}

Offset phase locking controls slave laser $v$ relative to reference laser $u$. The phasemeter tracks the detected beatnote phase through a digital phase-locked loop \cite{Gerberding2013Phasemeter}. Laser feedback uses the resulting measurement to control the slave laser \cite{BayleHartwig2023Unified}. The reference contribution to the phase error is represented by
\begin{equation}
 e_{uv}(t)=\phi_v(t)-\mathcal D_{uv}\phi_u(t)-\psi_{uv}(t).
 \label{eq:locking_error}
\end{equation}
The reference delay $\mathcal D_{uv}$ includes propagation and the optical and electronic delays along the reference path. The prescribed phase $\psi_{uv}$ combines the frequency offset, command modulation, and known carrier-propagation contribution in the adopted phase convention. A constant angular-frequency offset contributes a phase linear in time. In the high-gain band,
\begin{equation}
 \phi_v\simeq\mathcal D_{uv}\phi_u+\psi_{uv}+\epsilon_{uv}.
 \label{eq:locked_phase}
\end{equation}
Substitution in Eq.~\eqref{eq:locking_error} gives $e_{uv}\simeq\epsilon_{uv}$. Thus $\epsilon_{uv}$ is the residual of the same closed-loop error, with finite-gain, sensing, actuation, and readout contributions. An edge $u\to v$ represents this reference dependency.

For an ordered path $a=v_0\to v_1\to\cdots\to v_s=v$, write $\mathcal D_h=\mathcal D_{v_{h-1}v_h}$, $\psi_h=\psi_{v_{h-1}v_h}$, and $\epsilon_h=\epsilon_{v_{h-1}v_h}$. These quantities refer to the reference link entering laser $v_h$. Repeated substitution gives
\begin{align}
 \phi_v&\simeq\mathcal D_s\cdots\mathcal D_1\phi_a\nonumber\\
 &\quad+\sum_{h=1}^{s}
 (\mathcal D_s\cdots\mathcal D_{h+1})(\psi_h+\epsilon_h).
 \label{eq:locking_path}
\end{align}
The ordered factors act on the prescribed phase and residual error as well as the primary phase. At $h=s$, the empty downstream product is the identity $I$. Locking transfers delayed phase fluctuations and introduces correlations that enter the TDI noise spectra \cite{Staab2024LaserResiduals}.

\subsection{Phase measurements and second-generation TDI}\label{subsec:tdi_principle}

Choose $C_1=\phi_{12}$, $C_2=\phi_{23}$, and $C_3=\phi_{31}$ as reference phases on the three spacecraft. The physical primary remains $L_{12}$. For the laser-phase reduction, each local comparison is represented by the ideal phase difference between the two lasers. With this convention, the reduced stream is
\begin{align}
 \eta_{ij}&=s_{ij}-\mathcal D_{ij}(\phi_{ij}-C_i)
                     +(\phi_\ell-C_j).
 \label{eq:raw_reduction}
\end{align}
Each phase difference on the right is zero for the selected reference laser and is represented by the corresponding ideal local phase difference for the other laser. The laser contribution to $\eta_{ij}$ is therefore $\mathcal D_{ij}C_i-C_j$. Local readout noise and auxiliary corrections enter the complete instrument model, which includes optical-bench, test-mass, clock, and modulation measurements \cite{BayleHartwig2023Unified}.

A TDI observable has the form
\begin{equation}
 \mathcal O=\sum_{i\ne j}\mathcal P_{ij}(\{\mathcal D\})\eta_{ij},
 \label{eq:general_tdi}
\end{equation}
where $\mathcal P_{ij}$ is the delay-polynomial coefficient of stream $\eta_{ij}$ for the selected observable. The coefficient of reference laser $C_k$ is
\begin{equation}
 \mathcal C_k=\sum_{i\ne j}\mathcal P_{ij}
 (\delta_{ik}\mathcal D_{ij}-\delta_{jk}),
 \label{eq:tdi_cancellation}
\end{equation}
with $\delta_{ik}$ the Kronecker delta. Commuting static delays permit exact cancellation. Under the slowly varying-arm approximation, second-generation Michelson combinations cancel the laser-noise residual terms linear in the arm-length rates \cite{Tinto2023SecondGen}.

For $X_2$ based at spacecraft 1, define
\begin{align}
 R_3&=\eta_{31}+\mathcal D_{31}\eta_{13},\nonumber\\
 R_2&=\eta_{21}+\mathcal D_{21}\eta_{12},\nonumber\\
 A&=\mathcal D_{31}\mathcal D_{13},\qquad
 B=\mathcal D_{21}\mathcal D_{12}.
 \label{eq:round_trips}
\end{align}
The laser contributions to $R_3$ and $R_2$ are $(A-I)C_1$ and $(B-I)C_1$, respectively. The adopted second-generation combination is
\begin{align}
 X_2&=(I-BA)(R_3+AR_2)\nonumber\\
 &\quad-(I-AB)(R_2+BR_3).
 \label{eq:X_tdi}
\end{align}
Its laser term is $(ABBA-BAAB)C_1=[AB,BA]C_1$, where $[U,V]=UV-VU$ denotes a commutator. Cyclic replacement gives $Y_2$ and $Z_2$. Exact integer-polynomial checks retain the ordered residual and verify its cancellation for commuting delays in both local-oscillator assignments. Section~\ref{sec:tdi} uses the nonzero coefficients to specify required optical inputs.

\section{Cascading robustness framework}\label{sec:framework}
\subsection{Model assumptions and functional support}\label{sec:mapping}

Thermal excursions, optical-power fluctuations, pointing disturbances, timing errors, and phase-tracking disturbances can affect reference transfer and phase measurement. Damage can prevent a resource from operating, and performance degradation can place its output outside the accepted range. Both reduce the support qualified for a specified task. We study this reduction through a statistical model with prescribed initial retention probabilities.

We use three terms throughout the paper. An architecture vertex identifies a physical function. A cascade layer is the modeled support network assigned to that vertex. A node is one equal-weight support unit within that layer. Each laser reference function has one associated layer in the locking calculation. Each spacecraft optical measurement function has one layer in the separate TDI-support calculation. The measurement model identifies the optical inputs and readout stages associated with these functions~\cite{BayleHartwig2023Unified}. Exchangeability of support units within a layer is an assumption of the statistical ensemble.

The macrostructure follows the selected physical architecture. For example, $L_{12}\to L_{13}$ identifies the reference used to control laser $L_{13}$. Its user layer represents support for that reference function, and the directed dependency specifies which modeled units require the provider layer. In the TDI calculation, the nonzero delay coefficients identify the optical inputs used by each receiving measurement function. The assigned dependency fractions then define its statistical support requirements. Sections~\ref{sec:locking} and \ref{sec:tdi} specify these two constructions separately.

The internal structure is an undirected ER network. An internal edge denotes an available support connection between two modeled units. Edges are assigned independently, and provider labels are independent of internal degree. The layer size $N_i$ and mean degree $k_i$ specify the network ensemble. This common connectivity model allows the comparison to isolate the interaction between local qualification, internal connectivity, and the architecture's directed dependencies.

A modeled unit contributes to final functional support when it retains its local qualification, satisfies its assigned provider requirements, and belongs to the layer's giant component. This giant-component criterion is the adopted functional rule of the percolation model~\cite{Gao2013Framework}. The resulting score measures connected functional support in the declared ensemble. The initial retention probabilities, dependency fractions, and mean degrees are control parameters for comparisons under that rule.

Let $V_i$ contain the $N_i$ support units in layer $i$. Here $i,j$ index layers, identified by laser labels in the locking results and spacecraft measurement labels in the TDI-support results. Each unit initially meets its local operating criterion with probability $p_i$, with every provider held nominal. Thus $p_i$ is the initial usable fraction, evaluated before dependency propagation and connectivity pruning. The scans vary $p_i$ directly and generate independent Bernoulli qualification flags with these probabilities. For laser layers, the two digits of the laser label form the subscript, so $p_{12}$ refers to laser $L_{12}$.

A fraction $q_{ji}$ of the modeled units in user layer $i$ requires support from provider layer $j$. The complementary fraction $1-q_{ji}$ retains the declared local support capacity independently of that provider. Autonomy concerns this support task, while optical reference transfer follows Sec.~\ref{subsec:phase_locking}. The matrix convention is $[Q]_{ij}=q_{ji}$, with users in rows and providers in columns. Arrows point from provider to user. The microscopic pairing on a reciprocal layer pair is inverse, so a reciprocal dependency returns to the same support unit. This is the no-feedback pairing condition of the adopted percolation framework~\cite{Gao2013Framework}.

\subsection{Connectivity response and cascade stages}\label{sec:stages}

Let $\pi_i(d)$ be the connectivity-degree distribution in layer $i$ and $z$ an auxiliary generating variable. The degree and excess-degree generating functions are \cite{Newman2001Generating}
\begin{equation}
 G_i(z)=\sum_{d=0}^{\infty}\pi_i(d)z^d,\qquad H_i(z)=\frac{G_i'(z)}{G_i'(1)}.
 \label{eq:generating}
\end{equation}
Following random retention of a fraction $x$, the fraction of retained nodes in the giant component satisfies \cite{Gao2013Framework}
\begin{equation}
 g_i(x)=1-G_i[1-x+xf_i],\qquad f_i=H_i[1-x+xf_i].
 \label{eq:general_g}
\end{equation}
Here $f_i=f_i(x)$ is the smallest solution of the second equation in $[0,1]$. The quantity $1-x+xf_i$ is the probability that following a connectivity edge does not lead to the giant component. For an undirected ER layer with mean degree $k_i$,
\begin{equation}
 g_i(x)=1-\exp[-k_i xg_i(x)].
 \label{eq:er}
\end{equation}
The physical solution is the largest nonnegative branch. It is zero for $k_i x\leq1$.

Dependency checks and connectivity pruning generate the successive cascade stages \cite{Buldyrev2010NatureCascade}. To specify the finite calculation unambiguously, let $D_{ji}\subseteq V_i$ be the user units requiring provider $j$ and let $m_{ji}$ map them to their paired provider units. Define
\begin{equation}
 \begin{aligned}
 U_{ji}(S)&=\{\nu\in V_i\mid\nu\notin D_{ji}\}\\
 &\quad\cup\{\nu\in D_{ji}\mid m_{ji}(\nu)\in S\}.
 \end{aligned}
 \label{eq:provider_set}
\end{equation}
Here $U_{ji}(S)$ contains user units with satisfied support when the provider set is $S$. Starting from the initial qualified sets $S_i^{(0)}$, a synchronous round is
\begin{align}
 T_i^{(t+1)}&=S_i^{(t)}\cap\bigcap_{j\in\mathcal I_i}U_{ji}(S_j^{(t)}),\nonumber\\
 S_i^{(t+1)}&=\operatorname{LC}_i(T_i^{(t+1)}).
 \label{eq:finite_stages}
\end{align}
$\mathcal I_i$ is the set of providers of layer $i$, and $\operatorname{LC}_i$ selects the largest connected component in the induced layer graph. Ties use the component with the smallest node index. Connectivity graphs, pairings, and the initial disturbance are fixed during a cascade. The nested sets stabilize after finitely many removals. Their fractions $\frac{|S_i^{(t)}|}{N_i}$ display the cascade process.

\subsection{Stationary equations and comparison scores}\label{sec:stationary}

The ensemble stationary equations use the effective usable fraction $x_i$ and the cavity fraction $y_{ji}$ of provider $j$, with its direct dependence on user $i$ excluded \cite{Gao2013Framework}. Define the provider factor, with the same provider-first index order as $q_{ji}$ and $y_{ji}$, by
\begin{equation}
 a_{ji}=1-q_{ji}+q_{ji}y_{ji}g_j(x_j).
 \label{eq:factor}
\end{equation}
The published stationary relations take the product form
\begin{align}
 x_i&=p_i\prod_{j\in\mathcal I_i}a_{ji},\nonumber\\
 y_{ij}&=p_i\prod_{\substack{h\in\mathcal I_i\\h\ne j}}a_{hi}.
 \label{eq:stationary}
\end{align}
The second line is evaluated for each outgoing edge $i\to j$. Excluding its reciprocal input accounts for the inverse node pairing. The final functional fraction in layer $i$ and the architecture mean are
\begin{equation}
 P_i=x_i g_i(x_i),\qquad
 \bar P=\frac{1}{n}\sum_{i=1}^{n}P_i,
 \label{eq:score}
\end{equation}
where $n$ is the number of functional layers. Throughout the paper, $P_i$ denotes the final functional fraction in layer $i$, and $\bar P$ is its equal-weight architecture mean. Finite cascade stages are identified explicitly by their round index.

The baseline fixes $k_i=8$ for every layer to compare architectures with a common ER connectivity response. The numerical tests include $k_i=6$ and 10 around the selected examples. Uniform disturbance sets $p_i=p$ to isolate the effect of reference depth. Localized disturbance changes one assigned fraction, and the heterogeneous grid changes two. On each listed edge the common dependency fraction is $q$. These parameter controls compare the same support ensemble across the candidate architectures.

At fixed $q$, the integrated comparison is
\begin{equation}
 \AUC(q)=\int_0^1\bar P(p,q)\,dp.
 \label{eq:auc}
\end{equation}
The area under the robustness curve summarizes the declared uniform-disturbance scan. A separate path baseline sets $g_i=1$ with the same graph, $p_i$, and $q$. It isolates the contribution of dependency propagation to the score.

\subsection{Numerical implementation and verification}\label{sec:numerics}

Bracketed scalar roots evaluate Eq.~\eqref{eq:er}, with $g_i=0$ imposed analytically at $k_i x_i\leq1$. Directed trees are evaluated in topological order. Homogeneous regular reciprocal graphs use the published scalar reduction \cite{Gao2013Framework}. Other graphs use monotone iteration from the maximum retained state. Both the stationary products and the scalar response equations are checked at the returned state. The default tolerance is $10^{-12}$, with 50000 iterations available for near-threshold checks. The stationary solver index labels numerical updates, while Eq.~\eqref{eq:finite_stages} labels finite cascade rounds.

The locking grid contains 41 values of $p$ and nine values of $q$ from 0.2 to 1. The support pattern grid contains 41 values of $p$ and twenty values of $q$ from 0 to 0.95. These give 6642 and 4100 records. Integrated scores use the composite trapezoidal rule on the same uniform $p$ grid, with spacing $\Delta p=\frac{1}{40}$. Refinement to 641 points changes the integrated locking scores by at most $2.20\times10^{-4}$. The three-candidate comparison uses 191 values of $p_{31}$ from 0.05 to 1 at fixed $q=0.9$. Separate display data resolve the continuous layer thresholds and discontinuous branches. Uniform locking curves use 2001 regular samples plus layer thresholds. Pairwise difference curves use 4001 regular dependency fractions plus all layer thresholds. Curves for the six-edge, B, and C patterns use 1001 regular samples plus their first and secondary transitions. A and D retain the 201-point display grid with their exact fold limits. Discontinuous transitions contribute two one-sided states at the same abscissa. Continuous onsets contribute one limiting state. Independent 60-digit calculations check the selected layer thresholds and scalar folds. Transition calculations also check the stationary equations and states on both sides of each event. Appendix~\ref{app:critical} defines these calculations.

Finite networks have 5000, 20000, and 80000 nodes per layer, with independent ER edges of probability $\frac{8}{N_i-1}$. These model sizes test the ensemble limit and finite-size effects. For each directed interlayer dependency $j\to i$, $\operatorname{round}(qN_i)$ user units are paired with distinct provider units in layer $j$. Reciprocal dependencies obey the no-feedback pairing condition stated above. The locking and reciprocal-pair suite uses 24 realizations per setting. The five-pattern suite uses twelve. An independent set-based implementation compares every node at every stage with the sparse-network implementation in 96 small-network tests. All 364 recorded stages agree exactly.

\begin{figure*}[t]
\centering
\includegraphics[width=\textwidth]{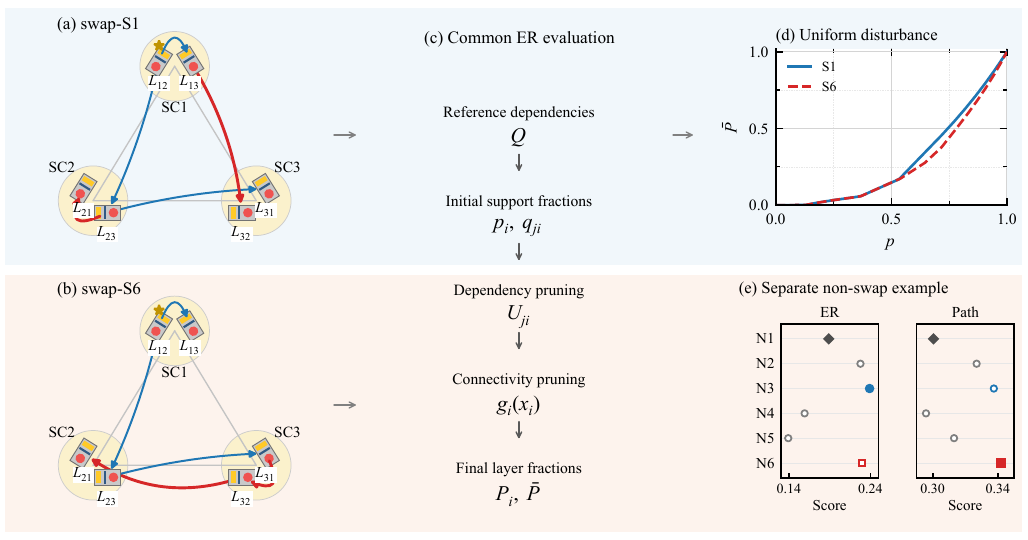}
\caption{Physical reference architectures and ER comparisons. Panels a and b show swap-S1 and swap-S6, with a star marking $L_{12}$. Blue and red arrows identify shared and changed reference edges, respectively. Panel c shows their common evaluation through dependency and connectivity pruning. Panel d compares uniform-disturbance scores at $q=1$. Panel e gives a separate six-candidate non-swap comparison under the disturbance of Fig.~\ref{fig:choices}. Its ER and path scores use separate scales. Gray diamonds mark catalogue choice N1. Filled blue and red markers identify the respective maxima N3 and N6. Every layer has $k_i=8$.}
\label{fig:overview}
\end{figure*}

\section{Phase-locking architectures and results}\label{sec:locking}
\subsection{Reference graphs and uniform disturbances}\label{sec:uniform}

The six laser reference functions form the vertex set $\{L_{12},L_{13},L_{21},L_{23},L_{31},L_{32}\}$. Each listed edge $u\to v$ supplies the phase reference for slave laser $v$. Table~\ref{tab:architectures} retains all eighteen catalogue configurations for primary $L_{12}$ \cite{Heinzel2025Catalog}. We checked all ninety reference edges against the local-oscillator assignments and the corresponding local and inter-spacecraft measurements. Figure~\ref{fig:overview} locates two of these reference trees within the spacecraft layout.

\begin{table*}[t]
\centering
\caption{All eighteen reference trees with primary $L_{12}$ from the locking catalogue \cite{Heinzel2025Catalog}. Edges point from reference provider to slave laser. Total depth is the sum of the six reference depths, and the local count gives the number of locks within one spacecraft.}
\label{tab:architectures}
\footnotesize\renewcommand{\arraystretch}{1.12}
\setlength{\tabcolsep}{3pt}
\begin{tabularx}{\textwidth}{@{}lccL@{}}
\toprule Configuration & Total depth & Local count & Reference edges \\
\midrule
non-swap-N1 & 9 & 3 & $L_{12}\to L_{13}$, $L_{12}\to L_{21}$, $L_{21}\to L_{23}$, $L_{13}\to L_{31}$, $L_{31}\to L_{32}$ \\
non-swap-N2 & 9 & 2 & $L_{12}\to L_{13}$, $L_{12}\to L_{21}$, $L_{21}\to L_{23}$, $L_{13}\to L_{31}$, $L_{23}\to L_{32}$ \\
non-swap-N3 & 11 & 3 & $L_{12}\to L_{13}$, $L_{12}\to L_{21}$, $L_{21}\to L_{23}$, $L_{32}\to L_{31}$, $L_{23}\to L_{32}$ \\
non-swap-N4 & 11 & 2 & $L_{12}\to L_{13}$, $L_{12}\to L_{21}$, $L_{32}\to L_{23}$, $L_{13}\to L_{31}$, $L_{31}\to L_{32}$ \\
non-swap-N5 & 15 & 3 & $L_{12}\to L_{13}$, $L_{23}\to L_{21}$, $L_{32}\to L_{23}$, $L_{13}\to L_{31}$, $L_{31}\to L_{32}$ \\
non-swap-N6 & 15 & 2 & $L_{31}\to L_{13}$, $L_{12}\to L_{21}$, $L_{21}\to L_{23}$, $L_{32}\to L_{31}$, $L_{23}\to L_{32}$ \\
\midrule
swap-S1 & 8 & 2 & $L_{12}\to L_{13}$, $L_{23}\to L_{21}$, $L_{12}\to L_{23}$, $L_{23}\to L_{31}$, $L_{13}\to L_{32}$ \\
swap-S2 & 9 & 3 & $L_{12}\to L_{13}$, $L_{23}\to L_{21}$, $L_{12}\to L_{23}$, $L_{32}\to L_{31}$, $L_{13}\to L_{32}$ \\
swap-S3 & 9 & 3 & $L_{12}\to L_{13}$, $L_{23}\to L_{21}$, $L_{12}\to L_{23}$, $L_{23}\to L_{31}$, $L_{31}\to L_{32}$ \\
swap-S4 & 9 & 1 & $L_{12}\to L_{13}$, $L_{32}\to L_{21}$, $L_{12}\to L_{23}$, $L_{23}\to L_{31}$, $L_{13}\to L_{32}$ \\
swap-S5 & 10 & 2 & $L_{12}\to L_{13}$, $L_{32}\to L_{21}$, $L_{12}\to L_{23}$, $L_{32}\to L_{31}$, $L_{13}\to L_{32}$ \\
swap-S6 & 11 & 2 & $L_{12}\to L_{13}$, $L_{32}\to L_{21}$, $L_{12}\to L_{23}$, $L_{23}\to L_{31}$, $L_{31}\to L_{32}$ \\
swap-S7 & 11 & 2 & $L_{21}\to L_{13}$, $L_{23}\to L_{21}$, $L_{12}\to L_{23}$, $L_{23}\to L_{31}$, $L_{31}\to L_{32}$ \\
swap-S8 & 12 & 1 & $L_{21}\to L_{13}$, $L_{23}\to L_{21}$, $L_{12}\to L_{23}$, $L_{23}\to L_{31}$, $L_{13}\to L_{32}$ \\
swap-S9 & 13 & 3 & $L_{12}\to L_{13}$, $L_{32}\to L_{21}$, $L_{21}\to L_{23}$, $L_{32}\to L_{31}$, $L_{13}\to L_{32}$ \\
swap-S10 & 15 & 2 & $L_{12}\to L_{13}$, $L_{32}\to L_{21}$, $L_{21}\to L_{23}$, $L_{23}\to L_{31}$, $L_{13}\to L_{32}$ \\
swap-S11 & 15 & 2 & $L_{21}\to L_{13}$, $L_{23}\to L_{21}$, $L_{12}\to L_{23}$, $L_{32}\to L_{31}$, $L_{13}\to L_{32}$ \\
swap-S12 & 15 & 1 & $L_{21}\to L_{13}$, $L_{32}\to L_{21}$, $L_{12}\to L_{23}$, $L_{23}\to L_{31}$, $L_{31}\to L_{32}$ \\
\botrule
\end{tabularx}
\end{table*}

For a tree, each nonprimary layer has one provider. Under common $p$, $q$, and $g$, the final fraction depends on the reference depth $d$,
\begin{align}
 P_0&=p\,g(p),\nonumber\\*
 x_d&=p[1-q+qP_{d-1}],\qquad P_d=x_dg(x_d),\nonumber\\*
 \bar P&=\frac16\sum_d n_dP_d.
 \label{eq:depth}
\end{align}
Here $n_d$ counts lasers at depth $d$, including the primary at depth zero. Equal depth distributions therefore give equal uniform-disturbance curves. In the path baseline at $q=1$, $P_d=p^{d+1}$. These relations identify the part of the ordering already explained by reference depth.

\begin{figure}[t]
\centering\includegraphics[width=\columnwidth]{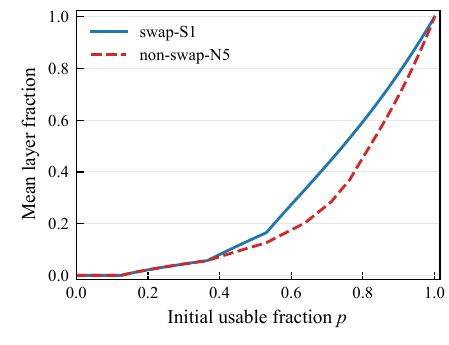}
\caption{Uniform-disturbance curves for the representative reference trees at $q=1$ and $k_i=8$. The final score averages the six layer fractions. S1 and N5 belong to different optical layouts and illustrate the dependence on reference depth. Selection tests compare configurations within each layout.}
\label{fig:uniform}
\end{figure}

\begin{figure}[t]
\centering\includegraphics[width=\columnwidth]{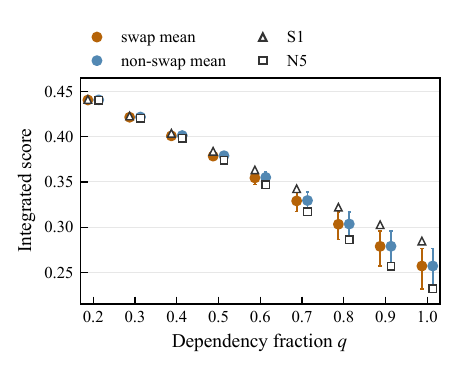}
\caption{Integrated uniform-disturbance scores. Orange and blue dots give swap and non-swap means at all nine dependency fractions. Capped intervals give the 15th to 85th configuration percentiles. Triangles and squares identify S1 and N5. Family positions are horizontally offset for legibility. Integration uses the same 41-point $p$ grid.}
\label{fig:auc}
\end{figure}

Figure~\ref{fig:uniform} shows the representative S1 and N5 comparison. Figure~\ref{fig:auc} integrates the same score for all catalogue configurations. Family means and percentile intervals across configurations summarize the declared ensemble. In all 738 uniform-disturbance family comparisons, the ER and path optimizing sets intersect. This uniform calculation establishes the reference case for the heterogeneous tests.

The ER response also supplies a connected support requirement. Define
\begin{equation}
 p_{\rm all}(q)=\inf\{p\mid\min_iP_i(p,q)>0\}.
 \label{eq:all_threshold}
\end{equation}
For $q=1$ and $k_i=8$, this threshold is 0.5309793 for S1 and 0.7635929 for N5. It is found from $\min_i x_i=\frac18$. The path baseline has positive fractions for every $p>0$.

\subsection{Nonuniform disturbances and architecture choice}\label{sec:heterogeneous}

The heterogeneous scan assigns fractions from $\{0.15,0.25,0.40,0.55,0.70,0.85\}$ to every pair of laser layers, sets the other four fractions to one, and uses $q\in\{0.6,0.9,1\}$. All fifteen pairs and all fraction combinations are retained. The resulting 1620 scenarios give 3240 decisions across the two hardware families. A maximizing set contains every configuration within $10^{-11}$ of the largest score. The ER and path maximizing sets are disjoint in 54 decisions and overlap in the other 3186. The control with one disturbed layer varies each of the six layers over the same six initial fractions and three dependency fractions. Comparing the two hardware families separately gives 216 decisions, of which two have disjoint maximizing sets.

Grouping by dependency fraction gives 4 disjoint decisions at $q=0.6$ and 50 at $q=0.9$, out of 1080 comparisons at each value. At $q=1$, the maximizing sets intersect in all 1080 comparisons. Fifty of the 54 disjoint decisions include an initial disturbance of the primary reference layer. The other four retain its initial fraction at one. Thus the observed change in the selected architecture occurs within the tested ensembles with partial dependence and is concentrated in disturbances involving the primary reference support.

The matched N1/N2 pair has the same complete depth distribution. Setting only $p_{31}=0.8$ with $q=1$ gives ER means 0.9320282 and 0.9656779, respectively. Disturbing $L_{23}$ instead interchanges these values. The path baseline gives the same preferences. This control isolates the role of disturbance location before the comparison of connected support responses.

Consider the non-swap family with $p_{12}=0.25$, $p_{31}=0.55$, unit fractions elsewhere, $q=0.9$, and $k_i=8$. Table~\ref{tab:criteria} places the three criteria alongside the complete comparison of all six candidates. The catalogue first minimizes total reference depth and then favors more local locks \cite{Heinzel2025Catalog}. It prefers N1. The path baseline selects N6. The ER score selects N3.

\begin{table}[t]
\centering
\caption{Complete non-swap comparison at $p_{12}=0.25$, $p_{31}=0.55$, unit fractions elsewhere, $q=0.9$, and $k_i=8$. The catalogue criteria prefer N1 through its depth and local count. All scores are evaluated on this same family.}
\label{tab:criteria}
\footnotesize\renewcommand{\arraystretch}{1.12}
\setlength{\tabcolsep}{3pt}
\begin{tabular}{@{}lcccc@{}}
\toprule Tree & Depth & Local & ER score & Path score \\
\midrule
N1 & 9 & 3 & 0.188791 & 0.300444 \\
N2 & 9 & 2 & 0.227801 & 0.326937 \\
N3 & 11 & 3 & 0.239040 & 0.337518 \\
N4 & 11 & 2 & 0.159154 & 0.295837 \\
N5 & 15 & 3 & 0.139044 & 0.313066 \\
N6 & 15 & 2 & 0.229851 & 0.341922 \\
\botrule
\end{tabular}
\end{table}

\begin{figure}[t]
\centering\includegraphics[width=\columnwidth]{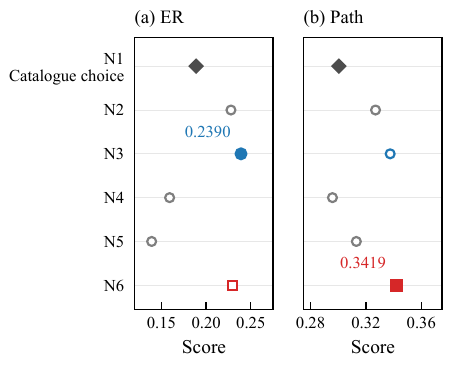}
\caption{Selection among all six non-swap candidates at $p_{12}=0.25$, $p_{31}=0.55$, unit fractions elsewhere, $q=0.9$, and $k_i=8$. Panels a and b give ER and path scores on separate horizontal scales. Filled blue and red markers identify the respective maxima N3 and N6. The gray diamond locates catalogue choice N1 in each panel. Table~\ref{tab:criteria} provides the scores.}
\label{fig:choices}
\end{figure}

The ER values are 0.1887905 for N1, 0.2390396 for N3, and 0.2298507 for N6. N3 exceeds the catalogue choice by 0.0502490 in mean layer fraction and the path choice by 0.0091888. The latter difference isolates the choice changed by including the connected support response under the same disturbance and dependency inputs. Figure~\ref{fig:choices} compares all six candidates at this input. Aligned rows identify the catalogue choice N1, the ER choice N3, and the path choice N6. Each panel ranks candidates within its own response model. Table~\ref{tab:criteria} gives the corresponding fixed-input scores.

N3 and N6 share the upstream chain ending at $L_{31}$. Only the reference supplying $L_{13}$ changes. With the nondecreasing response
\begin{equation}
 F_q(z)=[1-q+qz]g[1-q+qz],
 \label{eq:response_map}
\end{equation}
their difference is $\frac{F_q(P_{12})-F_q(P_{31})}{6}$. Under the path baseline, the corresponding provider fractions are 0.25 and 0.2793588, favoring N6. The ER values are 0.1992030 and 0.1529093, favoring N3. The reference graph is unchanged between the two calculations. The nonlinear response changes the ordering of the provider fractions.

A second example uses the swap family with $p_{13}=p_{23}=0.15$, unit fractions elsewhere, $q=0.9$, and $k_i=8$. S1 has ER score 0.1988159 and S6 has 0.1893596. The path scores are 0.3341667 and 0.3711417, respectively. The ER cascade along the longer S6 branch reaches $x_{21}<\frac18$, where the final giant component vanishes. The S1 branches retain positive fractions.

The comparison covers the complete grid specified above. The two representative cases illustrate the layer-level mechanisms of the changed preferences. Around each case, a neighborhood test gives each of the two assigned fractions its central value and offsets of $\pm0.02$, uses $k_i\in\{6,8,10\}$, and takes $q\in\{0.85,0.9,0.95\}$. The two cases, three values for each assigned fraction, three mean degrees and three dependency fractions give 162 family decisions. Fifty-four have disjoint optimizing sets. The comparison therefore has an explicit parameter domain and preserves the cases with coincident preferences.

\subsection{Finite network verification}\label{sec:finite}

The fixed-network suite contains 792 realizations of the single-layer reference, reciprocal-pair benchmarks, and the two selected locking comparisons. Initial masks and connectivity graphs are shared within a paired configuration comparison. Dependency matchings use fixed random seeds. At 80000 nodes per layer, S1 minus S6 has mean difference 0.0081490 with standard error 0.0004285. N3 minus N6 has mean difference 0.0093707 with standard error 0.0001434. All 24 paired realizations at this size have the sign predicted by the ensemble equations. The infinite-ensemble differences are 0.0094563 and 0.0091888.

\begin{figure}[t]
\centering\includegraphics[width=\columnwidth]{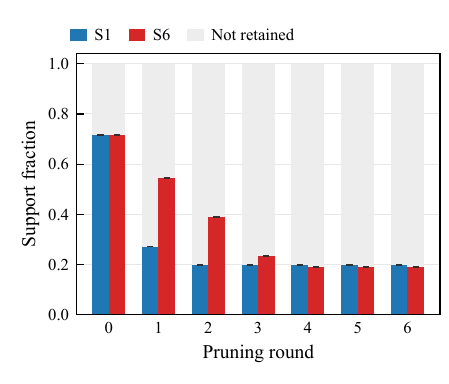}
\caption{Synchronous pruning for S1 and S6 with $p_{13}=p_{23}=0.15$, unit fractions elsewhere, $q=0.9$, $k_i=8$, and 80000 nodes per layer. Colored bars give means across 24 realizations with error bars spanning 1.96 standard errors. Gray portions include initial removal and subsequent pruning. Round zero contains the initial qualified sets. Later rounds follow Eq.~\eqref{eq:finite_stages}. Stable realizations are padded at their final values. Final means are 0.1982 for S1 and 0.1900 for S6.}
\label{fig:cascade_trace}
\end{figure}

Figure~\ref{fig:cascade_trace} follows the synchronous provider and connectivity checks of Eq.~\eqref{eq:finite_stages} on fixed graphs. S1 retains less connected support after the first round but more after convergence. Successive pruning rounds therefore reveal the emergence of the stationary preference.

Paired comparisons at three layer sizes assess the finite-size effects. At 80000 nodes, S1 minus S6 lies 3.05 standard errors below its ensemble limit. The predicted signs persist with finite-size bias retained.

\section{TDI optical-input support}\label{sec:tdi}
\subsection{Delay-polynomial support and graph direction}\label{sec:tdi_mapping}

The spacecraft measurement units $\mathrm{SC}_1$, $\mathrm{SC}_2$, and $\mathrm{SC}_3$ acquire the phase measurements entering Eq.~\eqref{eq:general_tdi}. A nonzero coefficient specifies an optical input used by the selected observable. Under the sender-first convention, $\eta_{ji}$ is formed at receiving unit $i$ using light from provider $j$. The support adjacency is therefore
\begin{equation}
 A_{ij}=
 \begin{cases}
 1,&\mathcal P_{ji}(\{\mathcal D\})\ne0,\\
 0,&\mathcal P_{ji}(\{\mathcal D\})=0.
 \end{cases}
 \label{eq:tdi_mapping}
\end{equation}
Rows are receiving users, columns are optical-input providers, and the arrow is $j\to i$. Each receiving measurement function includes its local acquisition resources and the required remote optical inputs. The support ensemble assigns a dependency fraction $q_{ji}=qA_{ij}$ to these functions. Thus the polynomial specifies the selected inputs and the declared pairing policy specifies resource dependence. The ER calculation evaluates the latter functional support model.

Table~\ref{tab:motifs} lists the four-edge patterns and the six-edge reference. There are fifteen labeled directed four-edge graphs on three vertices with distinct endpoints. Relabeling the spacecraft reduces them to four isomorphism classes. Common $p$, $q$, and $k_i$ give identical equations within a class. We label the four support patterns A through D and use 6L for the six-edge reference.

\begin{table*}[t]
\centering
\caption{Directed optical-input support patterns and their conditional ER thresholds at $q=0.95$ and $k_i=8$. The four-edge patterns form four isomorphism classes under relabeling of the three units. Labels A to D identify these patterns, and 6L denotes the six-edge reference. The onset threshold of the architecture mean is $p_c$, while $p_{\rm all}$ is the infimum for positive giant components in all three layers, as defined in Eq.~\eqref{eq:all_threshold}.}
\label{tab:motifs}
\footnotesize\renewcommand{\arraystretch}{1.12}
\setlength{\tabcolsep}{4pt}
\begin{tabularx}{\textwidth}{@{}lLcc@{}}
\toprule Pattern & Provider-to-user edges & $p_c$ & $p_{\rm all}$ \\
\midrule
TDI-6L & $\mathrm{SC}_{1}\to \mathrm{SC}_{2}$, $\mathrm{SC}_{2}\to \mathrm{SC}_{1}$, $\mathrm{SC}_{1}\to \mathrm{SC}_{3}$, $\mathrm{SC}_{3}\to \mathrm{SC}_{1}$, $\mathrm{SC}_{2}\to \mathrm{SC}_{3}$, $\mathrm{SC}_{3}\to \mathrm{SC}_{2}$ & 0.9931908 & 0.9931908 \\
TDI-A & $\mathrm{SC}_{1}\to \mathrm{SC}_{2}$, $\mathrm{SC}_{2}\to \mathrm{SC}_{1}$, $\mathrm{SC}_{1}\to \mathrm{SC}_{3}$, $\mathrm{SC}_{3}\to \mathrm{SC}_{1}$ & 0.7040166 & 0.7040166 \\
TDI-B & $\mathrm{SC}_{1}\to \mathrm{SC}_{2}$, $\mathrm{SC}_{2}\to \mathrm{SC}_{1}$, $\mathrm{SC}_{1}\to \mathrm{SC}_{3}$, $\mathrm{SC}_{2}\to \mathrm{SC}_{3}$ & 0.5320249 & 0.6542540 \\
TDI-C & $\mathrm{SC}_{1}\to \mathrm{SC}_{2}$, $\mathrm{SC}_{1}\to \mathrm{SC}_{3}$, $\mathrm{SC}_{2}\to \mathrm{SC}_{3}$, $\mathrm{SC}_{3}\to \mathrm{SC}_{2}$ & 0.1250000 & 0.7235849 \\
TDI-D & $\mathrm{SC}_{1}\to \mathrm{SC}_{2}$, $\mathrm{SC}_{2}\to \mathrm{SC}_{1}$, $\mathrm{SC}_{2}\to \mathrm{SC}_{3}$, $\mathrm{SC}_{3}\to \mathrm{SC}_{1}$ & 0.9707416 & 0.9707416 \\
\botrule
\end{tabularx}
\end{table*}

An exact polynomial construction checks that every listed four-edge pattern can support a static combination that cancels laser noise. Form the $3\times4$ laser-coefficient matrix with column $ji$ equal to $\delta_{jk}\mathcal D_{ji}-\delta_{ik}$ in row $k$. Its signed $3\times3$ minors give a polynomial null vector for commuting delays. All four coefficients are nonzero for each listed pattern, and the laser residual vanishes exactly. The ordered second-generation Michelson algebra is checked separately as described in Sec.~\ref{subsec:tdi_principle}. These tests verify the input support and delay conventions used in the mapping.

\subsection{Conditional support curves and transitions}\label{sec:tdi_results}

\begin{figure}[t]
\centering\includegraphics[width=\columnwidth]{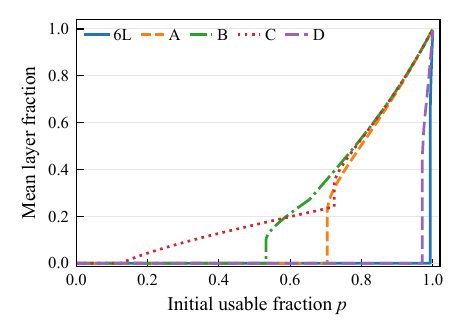}
\caption{Conditional ER scores at $q=0.95$ for the five patterns in Table~\ref{tab:motifs}, labeled 6L and A to D. The ordinate averages three functional giant-component fractions. Vertical segments connect one-sided limits at discontinuities, including the secondary receiving-pair onset of C. The downstream onset of B is continuous. Section~\ref{sec:tdi_mapping} defines the optical-input support policy.}
\label{fig:motif_curves}
\end{figure}

\begin{figure*}[t]
\centering\includegraphics[width=0.94\textwidth]{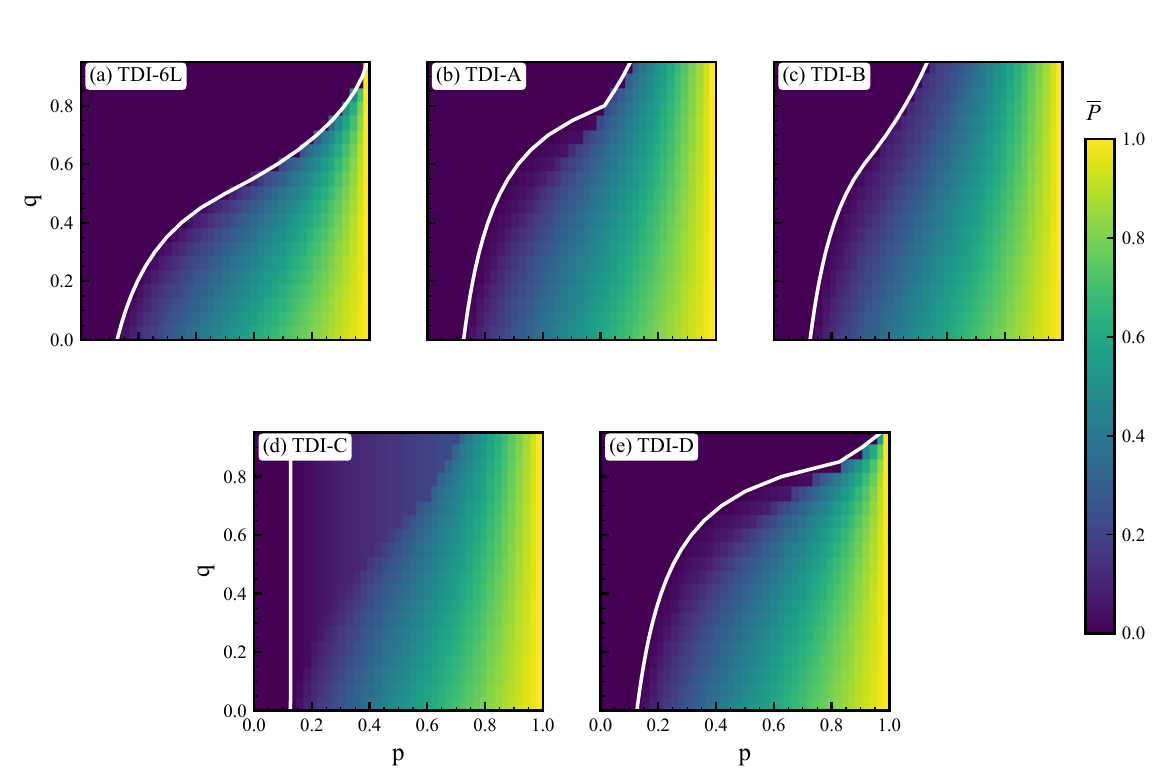}
\caption{Support-pattern landscapes on a common color scale. Colors give $\bar P$. White lines locate the display threshold $p_*(q)$ at $\bar P\ge10^{-3}$. Values at the twenty displayed dependency fractions follow the continuous crossing, with the fold used when a jump crosses the level. Adjacent points are joined for display. Appendix~\ref{app:critical} specifies the threshold calculations.}
\label{fig:landscapes}
\end{figure*}

Figure~\ref{fig:motif_curves} compares the declared support ensembles at $q=0.95$ and $k_i=8$. This fraction probes strong partial dependence while keeping a positive autonomous fraction. The full landscape scans $0\leq q\leq0.95$ in Fig.~\ref{fig:landscapes}. Patterns A through D have four edges each, so their different curves arise from the placement of reciprocal support and directed paths under the same edge count and common local response.

The onset threshold of the architecture mean is $p_c=\inf\{p\mid\bar P(p,q)>0\}$. The independent source in pattern C has $P_1=p\,g(p)$, giving a continuous onset at $p_c=\frac18$ and a positive mean for $p>p_c$. Pattern B contains an upstream reciprocal pair and a downstream receiving unit. Its first onset follows the reciprocal-pair threshold. Patterns A and D can develop positive branch folds at stronger dependency fractions. At $q=0.95$, the onset threshold of the six-edge reference is 0.9931908. Table~\ref{tab:motifs} gives the onset threshold and the all-layer threshold of Eq.~\eqref{eq:all_threshold} for every pattern.

The two criteria distinguish source support from connected support in every layer. At $p=0.2$ and $q=0.95$, pattern C has $P_1=0.1283963$ and $P_2=P_3=0$. Its mean is positive because the source remains connected. All three layer fractions become positive above $p_{\rm all}=0.7235849$. At this secondary transition the receiving pair appears discontinuously, raising the mean from 0.2404430 to 0.3360293 while the source fraction remains continuous. Pattern B reaches positive support in every layer continuously at 0.6542540, after its discontinuous upstream-pair onset at 0.5320249. B has the lowest all-layer threshold, about 6.93 percentage points below C, whereas C has the earliest mean onset. The preferred support pattern therefore changes with the required extent of connected support. Appendix~\ref{app:critical} gives the reductions and independent root checks.

Here $p$ measures initial local qualification, while $\bar P$ measures final connected support. The level $\bar P=10^{-3}$ represents 0.1 percent of the original modeled support, averaged equally over the three layers. It fixes the common display convention for the white lines in Fig.~\ref{fig:landscapes}.

Fixed finite network tests cover all five patterns at three dependency fractions, three initial fractions, and three layer sizes, with twelve realizations per setting. The resulting 1620 realizations retain finite-size fluctuations around the transition. Near the fold of pattern A, $q=0.95$ and $p=0.7$ lies just below the ensemble onset. A surviving finite realization contributes a positive empirical mean at 80000 nodes, with a standard error of comparable size. Subcritical finite components likewise contribute fractions that depend on layer size.

The model comparisons favor arrangements whose required support remains connected under the assigned disturbance. Their interpretation follows the stated resource dependency policy, optical-input selection, and layer-mean objective. The full delay polynomials retain the observable-specific signal response and noise-cancellation conditions.

\section{Discussion}\label{sec:discussion}

The percolation calculation adds two kinds of information to a reference-path description. The connected support condition supplies an explicit threshold, and the nonlinear response can change the configuration maximizing the score under nonuniform disturbance. The uniform-disturbance calculation identifies a useful exact depth-distribution limit. The heterogeneous comparison then shows where the same path baseline and the ER response choose different candidates.

The non-swap example also places the result in direct relation to the published catalogue. The catalogue favors N1 through its total depth and local-lock criteria. Under the declared disturbance, the ER score favors N3 and the path baseline favors N6. Their distinct choices follow from distinct evaluation criteria and assumptions.

In this statistical comparison, the initial usable fraction describes local qualification with providers held nominal, the dependency fraction specifies which modeled units require a given provider, and the mean degree controls the internal connectivity response. Their values define the ensemble being compared. The local support task fixes the meaning of provider independence. All observed disjoint choices in the main grid occur at partial dependency, and 50 of 54 involve initial disturbance of the primary reference support. These results connect the selection change to the interaction between nonuniform qualification, partial dependence, and internal connectivity. Instrument-specific applications can refine the support task and parameter inputs using corresponding operating requirements and local response data.

For TDI, the delay algebra identifies the required optical measurements and their transmitting and receiving units. The separate functional support policy supplies the dependencies used by the percolation equations. The mean fraction and the all-layer threshold distinguish two support objectives. Pattern C supplies the earliest positive mean through its source, whereas B supplies the earliest onset of positive giant components in every layer at the stated strong-dependency input. Actual observables retain their ordered delays, multiplicities, signal responses, and noise terms. Each of these quantities enters its corresponding physical performance calculation.

The finite simulator and stationary solver provide complementary checks within the adopted ensemble. The first exposes successive reductions on fixed graphs, with direct node-by-node tests of the update rule. The second evaluates the large-ensemble fractions and their branches. Finite-size fluctuations remain visible near a transition. The analytic regular-network thresholds serve as implementation benchmarks, while the architecture choices under assigned disturbances are the model's selection results.

\section{Conclusion}\label{sec:conclusion}

We have applied cascade equations and percolation measures to rank phase-locking architectures by their retained connected functional support in specified ER ensembles. Under uniform disturbances, the score reduces to a function of the reference-depth distribution. Under nonuniform disturbances and partial dependence, the connectivity response can change the preferred reference provider and hence the selected configuration. In the non-swap example, the catalogue criteria prefer N1, the path baseline N6 and the ER criterion N3. The complete scan gives disjoint ER and path maximizing sets in 54 of 3240 comparisons made separately within the two hardware families, all at partial dependence. Every sampled full-dependency comparison shares at least one maximizing configuration. Paired finite-network calculations retain the predicted preference signs with the reported finite-size bias.

The phase-locking results favor reference arrangements that limit accumulated degradation and direct downstream functions to providers retaining stronger connected support. The separate TDI optical-input comparison shows that the required extent of support also changes the preference. At $q=0.95$ and $k_i=8$, C has the earliest positive mean, while B reaches connected support in every layer at an initial qualification threshold about 6.93 percentage points below C. Disturbance location, provider dependence and the required support therefore enter the architecture choice alongside reference depth. Within the stated statistical model, these robustness comparisons can complement phase-locking feasibility, signal response and noise performance when prioritizing configurations for space-based gravitational-wave detectors.

\appendix
\section{Reference-chain and reciprocal-pair reductions}\label{app:reductions}

For a one-way dependency $j\to i$ with $q_{ij}=0$, the cavity fraction is $y_{ji}=x_j$. Its factor is $1-q+qP_j$. Along a directed chain, let subscripts $h=0,\ldots,s$ label successive layers. The chain therefore obeys
\begin{equation}
 x_0=p_0,\qquad
 x_h=p_h[1-q+qP_{h-1}].
 \label{eq:chain}
\end{equation}
For a provider with several direct users, the same factor enters each user's equation with its assigned initial fraction. Iterating Eq.~\eqref{eq:chain} through each branch yields Eq.~\eqref{eq:depth} under uniform inputs.

At $q=1$ and $g=1$, the path score is
\begin{equation}
 \bar P(p)=\frac16\sum_d n_d p^{d+1}.
 \label{eq:path_limit}
\end{equation}
For S6 and S7 the difference is $\frac{p^2(1-p)^2(1+p)}{6}$, whose integral is $\frac{1}{120}$. Thus the full depth distribution resolves this equal-total-depth comparison in the path baseline.

For a reciprocal pair with common $q$, define $c_q=1-q$. Its provider factors satisfy
\begin{equation}
 a_{ji}=c_q+\frac{qP_j}{a_{ij}},\qquad
 a_{ij}=c_q+\frac{qP_i}{a_{ji}}.
 \label{eq:pair}
\end{equation}
For $q<1$, the positive solution is
\begin{align}
 a_{ji}&=\frac{c_q^2+q(P_j-P_i)+\sqrt{\Delta_{ij}}}{2c_q},\nonumber\\
 \Delta_{ij}&=c_q^4+2c_q^2q(P_i+P_j)\nonumber\\
 &\quad+q^2(P_i-P_j)^2.
 \label{eq:pair_closed}
\end{align}
This algebraic reduction provides an independent form for the fold calculations. Product cavities in Eq.~\eqref{eq:stationary} retain the endpoint cases.

\section{Critical branches and display levels}\label{app:critical}

For homogeneous ER layers with $m$ reciprocal providers, the reduction is~\cite{Gao2013Framework}
\begin{equation}
\begin{aligned}
 p(P)&=\frac{P}{[1-e^{-kP}]b_q(P)^m},\\
 b_q(P)&=\frac{1-q+\sqrt{(1-q)^2+4qP}}2.
\end{aligned}
 \label{eq:regular}
\end{equation}
Here $P$ is the common single-layer final fraction. The continuous threshold is $\frac{1}{k(1-q)^m}$ up to the transition boundary at $q=\frac{k+m-\sqrt{m^2+2km}}{k}$. On the discontinuous branch, the minimum of $p(P)$ gives the fold. For $k=8$, $m=2$, and $q=0.5$, the threshold is $p_c=0.5$. At this point $x_i=\frac18$ and the final giant components are zero.

For a general support pattern under uniform $p$, eliminate the cavity fractions using Eq.~\eqref{eq:pair_closed} and the one-way factors. Write
\begin{equation}
 F_i(\mathbf P,p)=p\prod_{j\in\mathcal I_i}a_{ji}(\mathbf P)
 \frac{1-e^{-k_iP_i}}{P_i}-1.
 \label{eq:fold_residual}
\end{equation}
$\mathbf P$ contains the layer final fractions. Equation~\eqref{eq:fold_residual} applies on branches with $P_i>0$ for the included layers. A positive branch fold satisfies $\mathbf F=\mathbf0$ and $\det\left[\frac{\partial F_i}{\partial P_j}\right]=0$. Continuous emergence of a vanishing component is determined from the corresponding fixed-point equations before division by $P_i$, with the already surviving layers kept on their stationary branch. The downstream onset in B is treated below. The computation retains the positive fold state, residual, singular-Jacobian check, and values on the largest physical branch on both sides.

The display threshold used by the white curves is
\begin{equation}
 p_*(q)=\inf\{p\mid\bar P(p,q)\ge10^{-3}\}.
 \label{eq:display_level}
\end{equation}
For a continuous crossing, a bracketed root finds $\bar P=10^{-3}$. When a jump crosses the level, $p_*$ is the fold location. At $q=0.5$ the symmetric reference has $P=\bar P$ and $p_*=0.5000046401614406$. Equation~\eqref{eq:regular} gives $p(P)=\frac12+\frac{14}{3}P^2+\mathcal O(P^3)$ near onset. The quadratic leading increment explains $p_*-p_c\simeq4.64\times10^{-6}$ at $P=10^{-3}$.

The all-layer condition in Eq.~\eqref{eq:all_threshold} distinguishes the downstream onset in B and the reciprocal-pair onset in C. Let $p_1(P)$ be the right side of Eq.~\eqref{eq:regular} with one reciprocal provider, so $m=1$, at $q=0.95$. In B the upstream pair has common fraction $P$, and the downstream effective fraction is $p_1(P)[1-q+qP]^2$. Its threshold solves $p_1(P)[1-q+qP]^2=\frac1k$. In C the independent source has fraction $P_0=p\,g(p)$, and the receiving pair has effective initial fraction $p[1-q+qP_0]$. The all-layer onset occurs when this effective fraction reaches the positive-branch minimum of $p_1(P)$. Independent 80-digit scalar calculations verify both roots, and stationary solutions at offsets of $10^{-4}$ check all three layer fractions on either side.

\section{Exact polynomial and stage checks}\label{app:verification}

Let $M$ be the $3\times4$ laser-coefficient matrix for a four-edge support pattern, with column $r$ corresponding to its sender-first stream. If $M_{\widehat r}$ deletes column $r$, the coefficients
\begin{equation}
 \mathcal P_r=(-1)^{r+1}\det M_{\widehat r},\qquad r=1,\ldots,4,
 \label{eq:cofactor}
\end{equation}
satisfy $M\boldsymbol{\mathcal P}=\mathbf0$ over the commuting delay ring by the determinant identity. Here $\boldsymbol{\mathcal P}$ collects the four polynomial coefficients. The integer-polynomial calculation verifies every coefficient and residual. The six-edge reference is also represented by the combined support of $X_2$, $Y_2$, and $Z_2$.

The independent stage test evaluates Eq.~\eqref{eq:finite_stages} in two ways. One implementation uses sparse Boolean masks and connected-component labels. The other uses Python sets, explicit provider tests, and graph traversal. They agree for every node at all 364 recorded stages in 96 tests with one to four layers. The implementation includes empty sets and reciprocal pairings.

\section*{Declarations}

\textbf{Acknowledgements}
This work was supported by the National Key R\&D Program of China under Grant No. 2022YFC2204602 and the National Natural Science Foundation of China under Grant No. 12405060.

\textbf{Correspondence}
Correspondence should be addressed to Pan-Pan Wang. E-mail ppwang@cqu.edu.cn.

\textbf{Funding}
National Key R\&D Program of China, Grant No. 2022YFC2204602 and National Natural Science Foundation of China, Grant No. 12405060.

\textbf{Data availability}
The numerical data supporting the results of this study are available from the corresponding author upon reasonable request.

\textbf{Use of artificial intelligence}
A large language model assisted with coding and writing, with the authors reviewing the content and taking responsibility for the final manuscript.

\textbf{Conflict of interest}
The authors declare no competing interests.

\bibliography{ref_EPJC}

@misc{AmaroSeoane2017LISA,
  author = {Amaro-Seoane, Pau and others},
  title = {{Laser Interferometer Space Antenna}},
  year = {2017},
  note = {arXiv:1702.00786}
}

@article{AmaroSeoane2023LISA,
  author = {Amaro-Seoane, Pau and Andrews, Jeff and Arca Sedda, Manuel and others},
  title = {{Astrophysics with the Laser Interferometer Space Antenna}},
  journal = {Living Rev. Relativ.},
  year = {2023},
  doi = {10.1007/s41114-022-00041-y},
  volume = {26},
  pages = {2}
}

@misc{Colpi2024LISADefinition,
  author = {Colpi, Monica and others},
  title = {{LISA Definition Study Report}},
  year = {2024},
  note = {arXiv:2402.07571}
}

@article{Luo2016TianQin,
  author = {Luo, Jun and Chen, Li-Sheng and Duan, Hui-Zong and others},
  title = {{TianQin: a space-borne gravitational wave detector}},
  journal = {Class. Quantum Grav.},
  year = {2016},
  doi = {10.1088/0264-9381/33/3/035010},
  volume = {33},
  pages = {035010}
}

@article{Luo2026TianQinWhitePaper,
  author = {Luo, Jun and An, Haipeng and Bian, Ligong and others},
  title = {{Fundamental physics and cosmology with TianQin}},
  journal = {Living Rev. Relativ.},
  year = {2026},
  doi = {10.1007/s41114-025-00063-2},
  volume = {29},
  pages = {1}
}

@article{HuWu2017Taiji,
  author = {Hu, Wen-Rui and Wu, Yue-Liang},
  title = {{The Taiji Program in Space for gravitational wave physics and the nature of gravity}},
  journal = {Natl. Sci. Rev.},
  year = {2017},
  doi = {10.1093/nsr/nwx116},
  volume = {4},
  pages = {685--686}
}

@article{Heinzel2024FrequencyPlanning,
  author = {Heinzel, Gerhard and {\'{A}}lvarez-Vizoso, Javier and Dovale-{\'{A}}lvarez, Miguel and Wiesner, Karsten},
  title = {{Frequency planning for LISA}},
  journal = {Phys. Rev. D},
  year = {2024},
  doi = {10.1103/physrevd.110.042002},
  volume = {110},
  pages = {042002}
}

@article{Wang2023GeometricTDISensitivity,
  author = {Wang, Pan-Pan and Qian, Wei-Liang and Wu, Zhang-Qi and Chen, Hao-Kang and Huang, Weisheng and Wu, Han-Zhong and Tan, Yu-Jie and Shao, Cheng-Gang},
  title = {{Sensitivity functions for geometric time-delay interferometry combinations}},
  journal = {Phys. Rev. D},
  year = {2023},
  doi = {10.1103/physrevd.108.044075},
  volume = {108},
  pages = {044075}
}

@article{Heinzel2025Catalog,
  author = {Heinzel, Gerhard and {\'{A}}lvarez-Vizoso, Javier and Dovale-{\'{A}}lvarez, Miguel},
  title = {{Complete catalog of laser locking configurations for LISA}},
  journal = {Phys. Rev. D},
  year = {2025},
  doi = {10.1103/4xbx-4dz4},
  volume = {112},
  pages = {062002}
}

@article{Wang2024MinimalNullTDI,
  author = {Wang, Gang},
  title = {{Time delay interferometry with minimal null frequencies}},
  journal = {Phys. Rev. D},
  year = {2024},
  doi = {10.1103/physrevd.110.042005},
  volume = {110},
  pages = {042005}
}

@article{Wang2024RobustNoiseTDI,
  author = {Wang, Gang},
  title = {{Enhancing noise characterization with robust time delay interferometry combination}},
  journal = {Phys. Rev. D},
  year = {2024},
  doi = {10.1103/physrevd.110.064085},
  volume = {110},
  pages = {064085}
}

@article{Chen2025TTLTDICombinations,
  author = {Chen, Hao-Kang and Wang, Pan-Pan and Shao, Cheng-Gang},
  title = {{Characteristics of tilt-to-length coupling coefficients for different time-delay interferometry combinations for LISA}},
  journal = {Class. Quantum Grav.},
  year = {2025},
  doi = {10.1088/1361-6382/ae008b},
  volume = {42},
  pages = {185008}
}

@article{Estabrook2000TDI,
  author = {Estabrook, F. B. and Tinto, Massimo and Armstrong, J. W.},
  title = {{Time-delay analysis of LISA gravitational wave data: Elimination of spacecraft motion effects}},
  journal = {Phys. Rev. D},
  year = {2000},
  doi = {10.1103/physrevd.62.042002},
  volume = {62},
  pages = {042002}
}

@article{Tinto2002TDI,
  author = {Tinto, Massimo and Estabrook, F. B. and Armstrong, J. W.},
  title = {{Time-delay interferometry for LISA}},
  journal = {Phys. Rev. D},
  year = {2002},
  doi = {10.1103/physrevd.65.082003},
  volume = {65},
  pages = {082003}
}

@article{TintoShaddock2003Impl,
  author = {Tinto, Massimo and Shaddock, Daniel A. and Sylvestre, Julien and Armstrong, J. W.},
  title = {{Implementation of time-delay interferometry for LISA}},
  journal = {Phys. Rev. D},
  year = {2003},
  doi = {10.1103/physrevd.67.122003},
  volume = {67},
  pages = {122003}
}

@article{Tinto2004Moving,
  author = {Tinto, Massimo and Estabrook, F. B. and Armstrong, J. W.},
  title = {{Time delay interferometry with moving spacecraft arrays}},
  journal = {Phys. Rev. D},
  year = {2004},
  doi = {10.1103/physrevd.69.082001},
  volume = {69},
  pages = {082001}
}

@article{TintoDhurandhar2021Review,
  author = {Tinto, Massimo and Dhurandhar, Sanjeev V.},
  title = {{Time-delay interferometry}},
  journal = {Living Rev. Relativ.},
  year = {2021},
  doi = {10.1007/s41114-020-00029-6},
  volume = {24},
  pages = {1}
}

@article{BayleHartwig2023Unified,
  author = {Bayle, Jean-Baptiste and Hartwig, Olaf},
  title = {{Unified model for the LISA measurements and instrument simulations}},
  journal = {Phys. Rev. D},
  year = {2023},
  doi = {10.1103/physrevd.107.083019},
  volume = {107},
  pages = {083019}
}

@article{Tinto2023SecondGen,
  author = {Tinto, Massimo and Dhurandhar, Sanjeev and Malakar, Dishari},
  title = {{Second-generation time-delay interferometry}},
  journal = {Phys. Rev. D},
  year = {2023},
  doi = {10.1103/physrevd.107.082001},
  volume = {107},
  pages = {082001}
}

@article{Staab2024LaserResiduals,
  author = {Staab, Martin and Lilley, Marc and Bayle, Jean-Baptiste and Hartwig, Olaf},
  title = {{Laser noise residuals in LISA from on-board processing and time-delay interferometry}},
  journal = {Phys. Rev. D},
  year = {2024},
  doi = {10.1103/physrevd.109.043040},
  volume = {109},
  pages = {043040}
}

@article{JiangZhang2026EPJCPolarization,
  author = {Jiang, Tong and Zhang, Chunyu},
  title = {{Constraint on gravitational-wave polarizations for space-based detectors with time-delay interferometry}},
  journal = {Eur. Phys. J. C},
  year = {2026},
  doi = {10.1140/epjc/s10052-026-15444-2},
  volume = {86},
  pages = {194}
}

@article{Zhang2026EPJCSensitivity,
  author = {Zhang, Chunyu},
  title = {{Analytical sensitivity curves for additional gravitational-wave polarizations of second-generation time-delay interferometry}},
  journal = {Eur. Phys. J. C},
  year = {2026},
  doi = {10.1140/epjc/s10052-026-15433-5},
  volume = {86},
  pages = {219}
}

@article{Liu2026EPJCTDICombinations,
  author = {Liu, Yong-Yong and Zhang, Jing-Rui and Du, Ming-Hui and Liu, He-Shan and Xu, Peng and Zhang, Yun-Long},
  title = {{Detectability of axion-like dark matter for different time-delay interferometry combinations in space-based gravitational wave detectors}},
  journal = {Eur. Phys. J. C},
  year = {2026},
  doi = {10.1140/epjc/s10052-026-15578-3},
  volume = {86},
  pages = {347}
}

@article{Buldyrev2010NatureCascade,
  author = {Buldyrev, Sergey V. and Parshani, Roni and Paul, Gerald and Stanley, H. Eugene and Havlin, Shlomo},
  title = {{Catastrophic cascade of failures in interdependent networks}},
  journal = {Nature},
  year = {2010},
  doi = {10.1038/nature08932},
  volume = {464},
  pages = {1025--1028}
}

@article{Parshani2010CouplingPRL,
  author = {Parshani, Roni and Buldyrev, Sergey V. and Havlin, Shlomo},
  title = {{Interdependent Networks: Reducing the Coupling Strength Leads to a Change from a First to Second Order Percolation Transition}},
  journal = {Phys. Rev. Lett.},
  year = {2010},
  doi = {10.1103/physrevlett.105.048701},
  volume = {105},
  pages = {048701}
}

@article{Gerberding2013Phasemeter,
  author = {Gerberding, Oliver and Sheard, Benjamin and Bykov, Iouri and Kullmann, Joachim and Delgado, Juan Jose Esteban and Danzmann, Karsten and Heinzel, Gerhard},
  title = {{Phasemeter core for intersatellite laser heterodyne interferometry: modelling, simulations and experiments}},
  journal = {Class. Quantum Grav.},
  year = {2013},
  doi = {10.1088/0264-9381/30/23/235029},
  volume = {30},
  pages = {235029}
}

@article{Gao2013Framework,
  author = {Gao, Jianxi and Buldyrev, Sergey V. and Stanley, H. Eugene and Xu, Xiaoming and Havlin, Shlomo},
  title = {{Percolation of a general network of networks}},
  journal = {Phys. Rev. E},
  year = {2013},
  doi = {10.1103/physreve.88.062816},
  volume = {88},
  pages = {062816}
}

@article{Newman2001Generating,
  author = {Newman, M. E. J. and Strogatz, S. H. and Watts, D. J.},
  title = {{Random graphs with arbitrary degree distributions and their applications}},
  journal = {Phys. Rev. E},
  year = {2001},
  doi = {10.1103/physreve.64.026118},
  volume = {64},
  pages = {026118}
}

\end{document}